\documentclass[aps,superscriptaddress,twocolumn,floatfix,prb]{revtex4-1}
\usepackage{graphicx}
\usepackage{color}
\usepackage{amsmath,amssymb}

\begin{document}

\title[Stability studies in trapless pBEC]{Stability Window of Trapless Polariton Bose-Einstein condensates}

\author{S. Sabari}
\email{ssabari01@gmail.com}
\affiliation{Institute of Theoretical Physics, UNESP -- Universidade Estadual Paulista, 01156-970 S\~{a}o Paulo, SP, Brazil.}
\affiliation{Centre for Nonlinear Science (CeNSc), Department of Physics, Government College for Women (A), Kumbakonam 612001, India.}

\author{R. Kishor Kumar}
\email{kishor.bec@gmail.com}
\affiliation{Department of Physics, Centre for Quantum Science, and Dodd-Walls Centre for Photonic and Quantum Technologies, University of Otago, Dunedin 9054, New Zealand.}

\author{R. Radha}
\email{vittal.cnls01@gmail.com}
\affiliation{Centre for Nonlinear Science (CeNSc), Department of Physics, Government College for Women (A), Kumbakonam 612001, India.}

\author{P. Muruganandam}
\email{anand@bdu.ac.in}
\affiliation{Department of Physics, Bharathidasan University, Tiruchirappalli 620024, India.}

\begin{abstract}
We theoretically explore the possibility of stabilizing the trapless polariton Bose-Einstein condensates (pBECs). Exploiting the variational method, we solve the associated nonlinear, complex Gross-Pitaevskii (cGP) equation and derive the equation of motion for the amplitude and width of the condensate. These variational results described by ordinary differential equations are rewritten to perform a linear stability analysis to generate a stability window in the repulsive domain. A set of coupled nonlinear ordinary differential equations obtained through variational approach are then solved by numerical simulations through the fourth order Runge-Kutta method, which are further supported by split-step Crank-Nicholson method, thereby setting the platform for stable pBECs. In particular, we generate a window containing system parameters in the $g_1-\gamma_{eff}$ space within which the system can admit stable condensates. The highlight of the results is that one observes beating effects in the real time evolution of the condensates with attractive interactions much similar to multicomponent BECs, and their periodicity can be varied by manipulating linear and nonlinear loss/gain terms. For repulsive condensates, one notices the stretching of the density. 
\end{abstract}

\date{\today}

\maketitle

\section{Introduction}\label{sec1} 

The recent upsurge in the field of nanophotonics devices and optical computing can be attributed to the strong interaction between photons and excitons, the composite particle being commonly known as Polaritons~\cite{Balili2007}. Excitons are the bound states of electron and hole pairs via Coulombian interaction. Polaritons are essentially hybrid light-matter-quasi particles resulting from the strong coupling between excitons and photons~\cite{Byrnes2014, Plumhof2014, Christopoulos2007, Balili2007, Carusotto2013}. This strong coupling leads to nonlinear interaction between them, thereby creating a unique laboratory for the investigation of nonlinear collective phenomena like Bose-Einstein condensation~\cite{Kasprzak2006, Balili2007, Carusotto2013}, superfluidity~\cite{Amo2009}, quantized vortices~\cite{Lagoudakis2008, Keeling2008} etc. The extremely small effective mass of the polaritons (essentially arising from the photonic part) of the order of $10^{-5}$ of the mass of free electrons and their large coherence length in the mm scale combined with the strong nonlinearity arising from the excitonic part results in a strong nonlinear optical response, which can have wider ramifications in optical switching~\cite{Jabri2020}, optical computing~\cite{Ghosh2020}, photonic neural networks~\cite{Xu2020} etc. Such quantum microcavities present a unique platform for the investigation of quantum collective phenomena like polariton Bose-Einstein condensates (pBECs) in nonequilibrium systems in the temperature regime ranging from a few kelvins to room temperature~\cite{Christopoulos2007}. The latter case may offer significant potential for practical applications as it enables one to realize a new source of coherent light and optical switching via the Kerr nonlinearity~\cite{Jabri2020}.

The pBECs are often described by the complex Gross-Pitaevskii (cGP) equation~\cite{Keeling2008}, which represents the non-equilibrium dynamics with pumping and decaying terms, unlike the conventional Gross-Pitaevskii (GP) equation widely used in the realm of equilibrium systems describing ultracold atoms. Considering the slow dynamics of the polariton condensates, one can eliminate the reservoir effects so that the complex-Gross-Pitaevskii equation reduces to a single partial differential equation. The above assumption can be vindicated by the fact that the lifetime of the polaritons is much larger than the cavity round trip time. In addition, the pumping threshold for condensates is much smaller than the threshold for photon lasing with electronic population inversion. 

The nonlinear nature and the interplay of the intrinsic nonlinearity due to inter-particle interactions and the loss/gain nature of the nonlinear interactions of the system had led to a wide variety of remarkable observations including, but certainly not limited to, solitons~\cite{Chen2018, Amo2011, Dominici2015, Sich2012}, bright solitons~\cite{Sich2012, Egorov2009}, dark solitons~\cite{Ma2017, Grosso2011}, gap solitons~\cite{Tanese2013, Ostrovskaya2013}, oblique dark solitons~\cite{El2006}, dissipative solitons~\cite{Ostrovskaya2012}, studies in spin-orbit coupled (SOC)-pBECs~\cite{Whittaker2018} and   dipolar-pBECs~\cite{Rosenberg2018}.  

Despite several experimental explorations on pBECs, the bottom line is that the pBECs are highly unstable and  stabilizing them continues to be challenging even today. Also, in recent times, the study of trapless systems has attracted considerable attention in many fields, including, nonlinear physics~\cite{Fetter2012}, optics~\cite{Towers2002, Fetter2012} and BECs~\cite{Saito2003, Adhikari2004, Abdullaev2003, Sabari2010, Sabari2015, Sabari2017}. Hence, in the present investigation, we plan to  study the stability properties of trapless pBECs in both attractive and repulsive regimes. The organization of the present paper is as follows. In section 2, we present the theoretical model that describes the pBECs. Then, we discuss the variational analysis of the problem and point out the possible means of stabilization of the pBECs in section 3 besides showing the stability domains. In section 4, we report the numerical results of the time-dependent GP equation through the split-step Crank-Nicholson (SSCN) method. Section 5 summarizes our results opening up new avenues in the investigation of pBECs.

\section{Theoretical model}\label{sec2}

The GP equation is a mean-field model developed to deal with many-body problems. This equation has a form similar to the nonlinear Schr\'odinger equation, which incorporates the external potential and inter-particle interaction properly. The non-resonantly pumped pBEC can be modelled with a generalized complex GP equation to describe the dynamics of the polariton condensates having the population of the ground state being consistently replenished from a reservoir of incoherent excitons. The classical field $\psi$, known as the macroscopic wavefunction or order parameter of the polariton condensate, is used to replace the field operator by ignoring the non-condensate part from quantum and thermal fluctuations. Under the assumption that the polariton gas is weakly-interacting at low-temperatures, the condensate dynamics can be described by the GP type equation without microscopic physics of the polaritons involved. Here, a theoretical model is presented for a pBEC as~\cite{Keeling2008}:
\begin{align}
 \mathrm{i} \hbar \frac{\partial}{\partial t}  \psi(r,t) = &\, \left[-\frac{\hbar^2}{2m_p}\nabla^2 + V_{3D} + \mathrm{i} \, \Gamma_{\text{eff}} \right] \psi(r,t)  \notag \\ 
 &\, + \left[\,g \vert \psi(r,t) \vert^2-i\,g_1 \vert \psi(r,t)\vert^2 \right] \psi(r,t), \label{equ1}
\end{align}
where the external trapping potential assumes the harmonic form  $V_{3D} = m_p \omega^2 r^2/2$, with $\omega$ being the angular frequency, $m_p$ the polariton mass, $r$ the radial coordinate. $g$ is the interaction energy and $\Gamma_{\text{eff}} (= \gamma - \kappa)$ is the linear net gain describing the balance of the stimulated scattering of polaritons into the condensate and the decay of polaritons out of the cavity. The above GP equation \eqref{equ1} is subjected to the following normalization condition:
\begin{align}
N = \int \lvert \psi \rvert^2 dr, \label{eq:norm}
\end{align}
where $N$ is the total number of polaritons, and $dr$ is the volume element. Due to the presence of gain and loss terms in Eq.~\eqref{equ1}, $N$ need not be constant.

It is more convenient to use the GP Eq.~(\ref{equ1}) in a dimensionless form. For this purpose, we shall make the transformation of variables as $\bar r = r/l$, $\bar t=t \omega$, and $\bar \psi=\psi /\sqrt{\hbar \omega}$, where the harmonic oscillator length $l=\sqrt{\hbar/(m_p \omega)}$. After removing the overbar, Eq.~(\ref{equ1}) can be rewritten as
\begin{align}
\mathrm{i} \frac{\partial}{\partial t}  \psi(r,t)  = &\, \left[-\frac{1}{2}\nabla^2+ \frac{\Omega^2 r^2}{2} + \mathrm{i} \, \gamma_{\text{eff}} \right] \psi(r,t)  \notag \\ 
&\, + \left[\,g \vert \psi(r,t) \vert^2-i\, g_1 \vert \psi(r,t)\vert^2 \right] \psi(r,t). \label{equ2}
\end{align} 
where $\gamma_{\text{eff}}=\Gamma_{\text{eff}}/\hbar \omega$. The range of the physical parameters provided based on the experiments of refs.~\cite{Kasprzak2006,Balili2007} are $\gamma_{\text{eff}}=0 \sim 20 \, \hbar\,\omega$, $g=0 \sim 50\, \hbar\,\omega$ and $g_1= 0 \sim 50\, \hbar\,\omega$.  

In the case of strong radial confinement, the motion of the atoms is free in the axial direction, but restricted in the radial direction. Under this condition, the dynamics of the pBECs can be effectively studied by the following one-dimensional GP equation
\begin{subequations}
\begin{align}
\left[\mathrm{i} \frac{\partial}{\partial t} +   \frac{1}{2}\frac{\partial^2}{\partial x^2} - V_{1D} - g \vert \psi(x,t) \vert^2 \right] \psi(x,t)
= W, \label{equ3}
\end{align}%
where
\begin{align}
W = \mathrm{i} \left[\gamma_{\text{eff}} - g_1 \vert \psi(x,t)\vert^2 \right] \psi(x,t), \label{equ3b}
\end{align}
\end{subequations}
and $V_{1D} = d(t) \Omega^2 x^2/2$ where $\Omega^2$ is the trap frequency and $d(t)$ represents the strength of the external trap which is to be reduced from $1$ to $0$ when the trap is switched off.

\section{Variational Results}
\label{sec3}

In order to obtain the governing equations of motion for the system parameters, we use the variational approach with the following Gaussian ansatz as a trial wave function.
\begin{align}
\psi(x,t) = A(t)\exp{\left[-\frac{x^2}{2R(t)^2} + \mathrm{i} \beta(t)x^2 + \mathrm{i} \alpha(t) \right]}, \label{equ4}
\end{align}
where $A(t)$, $R(t)$, $\beta(t)$ and $\alpha(t)$ are the time dependent amplitude, width, chirp and phase, respectively. Since the dissipative and amplifying terms have no  influence on the center-of-mass coordinate  in  the Gaussian ansatz, we ignore it in the present study. The variational approach is applied to the averaged Lagrangian of the conservative system 
\begin{align}\label{equ5}
L = \int {\cal L} \, dx ,
\end{align}
where the Lagrangian density ${\cal L}\equiv {\cal L} (x,t)$, is given by
\begin{align}
\mathcal{L}(t) = \frac{\mathrm{i}}{2} \left( \dot{\psi} \psi^* -\dot{\psi^*} \psi \right) - \frac{1}{2}\vert \nabla \psi\vert^2 - V_{1D}\vert\psi\vert^2 -\frac{1}{2}g \vert \psi\vert^4. \label{equ6} 
\end{align}
The trial wave function (\ref{equ4}) is substituted into (\ref{equ6}) and~(\ref{equ5}) to find the averaged Lagrangian in terms of the condensate wavefunction parameters as
\begin{align}
L = -\frac{\sqrt\pi }{2}&\,A(t)^2R(t) \left[\frac{1}{R(t)^2} + \frac{1}{\sqrt 2}\, g\, A(t)^2+ 2\, \Dot{\alpha} \right. \notag\\  &\, +\frac{d(t) \Omega^2}{2} R(t)^2+\left. R(t)^2 \left( \beta(t)^2 + \frac{\Dot{\beta}}{2} \right) \right]. \label{equ7}
\end{align} 
We formally add ${\cal L}_W$ to Eq.~(\ref{equ6})  with the property that $\delta {\cal L}_W/\delta \psi^*=-W$,  where $W$ is  given by Eq.~(\ref{equ3b}). By applying the Euler-Lagrange equations to  ${\cal L}'\equiv {\cal L} + {\cal L}_W$, with respect to $\psi^*$, we obtain 
\begin{align}\label{equ8}
\left[\frac{\partial {\cal L^\prime}}{\partial \psi^*} -
\frac{d}{dt}\frac{\partial{\cal L^\prime}}{\partial \Dot{\psi^*}}\right] 
= \left[\frac{\partial {\cal L}}{\partial \psi^*} -
\frac{d}{dt}\frac{\partial{\cal L}}{\partial \Dot{\psi^*}}\right] 
- W(\psi,\psi^*) = 0,
\end{align}
that leads to Eq.~(\ref{equ3}) (The conjugate equation is obtained
in a similar way).

The corresponding variational principle is given by
\begin{align}
\delta\int_0^t L'dt = \delta\int_0^t (L + L_W) dt = 0 ,
\end{align}
where, as in Eq.~(\ref{equ5}), $L_W = \int dx {\cal L}_W .$ 
Taking into account the fact that for a small shift $\delta\eta$ of some variational parameter $\eta$, we have
\begin{align}
f(\eta + \delta\eta) = f(\eta) + \delta\eta \frac{\partial f}{\partial\eta},
\end{align}
where $f\equiv f(\psi,\psi^\ast) = L$ or $L_W$, we obtain a system of equations for the variational parameters 
$\eta_i$: 
\begin{align}\label{equ11}
\frac{\partial L}{\partial \eta_i} -
\frac{d}{dt}\frac{\partial L}{\partial\dot\eta_{t}} = \int
d s \left[W\frac{\partial \psi^*}{\partial\eta_i} + W^*\frac{\partial
\psi}{\partial\eta_i}\right],
\end{align}
where \( \eta_i \in \left( A, R, \beta, \alpha \right)\).
By substituting Eqs.~(\ref{equ4}) and (\ref{equ7}) into Eq.~(\ref{equ11}), we obtain the following set of coupled nonlinear ordinary differential equations (ODEs)~\cite{Filho2001}
\begin{subequations}
\label{eq:odes:final}
\begin{align}
    \dot A & = (\gamma_{\text{eff}} - \beta) A - \frac{5  g_1}{4 \sqrt{2}} A^3
    , \label{eq:odes:final:a}\\
    \dot R & = 2 \beta R + \frac{g_1 }{2 \sqrt{2}} A^2 R
    , \label{eq:odes:final:b}\\
    \dot \beta & = -\frac{d(t)  \Omega^2}{2} + \frac{1}{2R^4} + \frac{g A^2}{2 \sqrt{2} R^2} - 2 \beta^2 
    , \label{eq:odes:final:c}\\
    \dot \alpha & = - \frac{1}{2R^2} - \frac{5 g }{4 \sqrt{2}}A^2 
\end{align}
Now, the normalization condition \eqref{eq:norm} leads to
\begin{align}
\dot N = 2 \sqrt{\pi} \gamma_{\text{eff}} A^2 R - \sqrt{2 \pi} g_1 A^4 R.
\end{align}
\end{subequations}
The above equation displays the distinct signature of nonequilibrium systems with reservoir terms where the number of atoms does not remain a constant. 

Multiplying (\ref{eq:odes:final:a}) by $2A$ and (\ref{eq:odes:final:b}) by $2R$, and redefining $A^2 = X$ and $R^2 = Y$, we get
\begin{subequations}
\label{eq:odes2:final}
\begin{align}
    \dot X & = 2 (\gamma_{\text{eff}} - \beta) X - \frac{5  g_1}{2 \sqrt{2}} X^2
    , \label{eq:odes2:final:a}\\
    \dot Y & = 4 \beta Y + \frac{g_1 }{\sqrt{2}} X Y
    , \label{eq:odes2:final:b}\\
    \dot \beta & = -\frac{d(t) \Omega^2}{2} + \frac{1}{2Y^2} + \frac{g}{2 \sqrt{2}} \frac{X}{Y} - 2 \beta^2 
    . \label{eq:odes2:final:c} 
\end{align}
\end{subequations}
A stationary solution to the above problem can be given as
\begin{subequations}
\begin{align}
X^* & = \frac{\sqrt{2} \gamma_{\text{eff}}}{g_1}, \\
Y^* & = \frac{2 g_0 \gamma_{\text{eff}} \pm
   2 \sqrt{g_1^2 \gamma _{\text{eff}}^2+g_0^2 \gamma
   _{\text{eff}}^2+4 g_1^2 \Omega}
   }{g_1 \gamma _{\text{eff}}^2+4 g_1 \Omega
   },  \\
\beta^* & =  -\frac{\gamma _{\text{eff}}}{4}.
\end{align}
\end{subequations} 
As $X$ and $Y$, respectively, represent to $A^2$ and $R^2$, both $X^*$ and $Y^*$ should be positive. We perform a linear stability analysis of the equilibrium points to identify the regions in the $\gamma_{\text{eff}} - g_1$ plane. %
\begin{figure}[!ht]
   \centering
   \includegraphics[width=\linewidth]{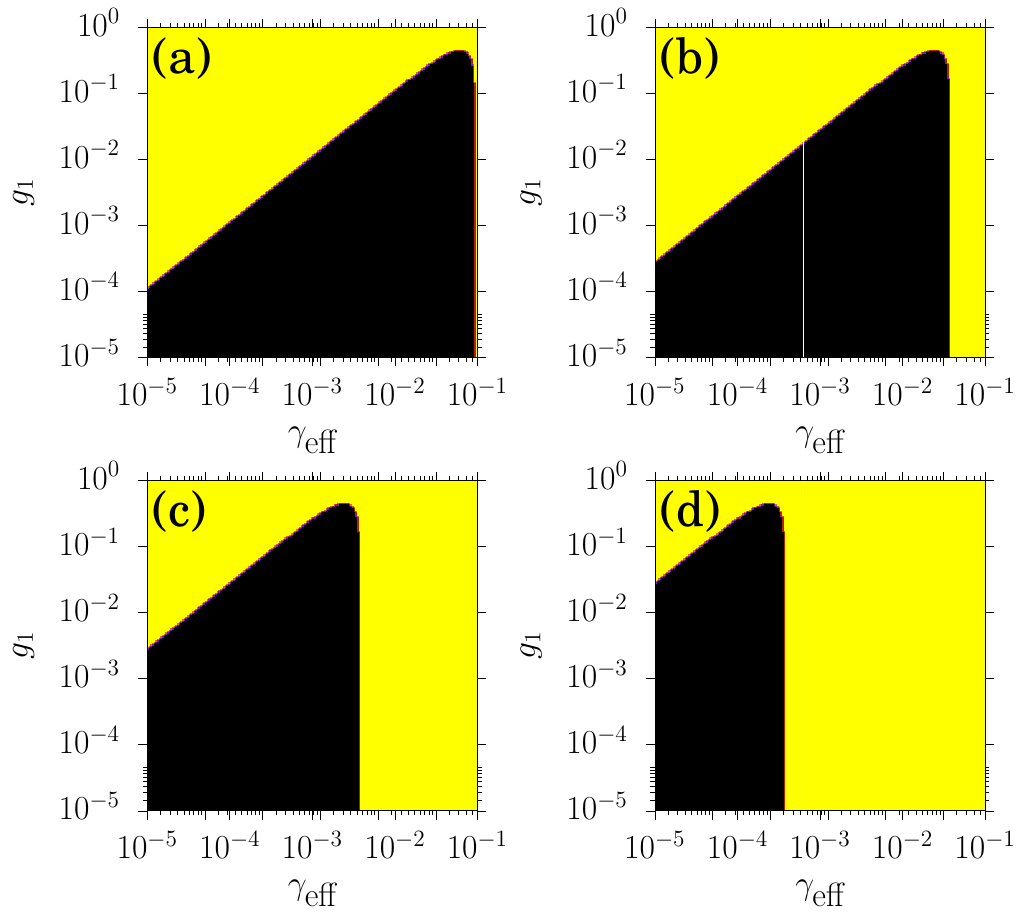}
   \caption{(Color online) Plot showing the stability regions (black) where the  equilibrium point $(X^*, Y^*, \beta^*)$ with $g = 5$ and $d(t) = 1$ is stable: (a) $\Omega = 0.25$, (b) $\Omega = 0.1$, (c) $\Omega = 0.01$, and (d)  $\Omega = 0.001$.}
   \label{fig:eig}
\end{figure}%
Fig.~\ref{fig:eig} illustrates the regions of stability of $(X^*, Y^*, \beta^*)$ (with $X^* > 0$ and $Y^* > 0$) in the $\gamma_{\text{eff}} - g_1 $ plane for a repulsive pBEC with $g = 5$ in the presence of trap with different strengths. %
\begin{figure}[!ht]
  \centering
  \includegraphics[width=\linewidth]{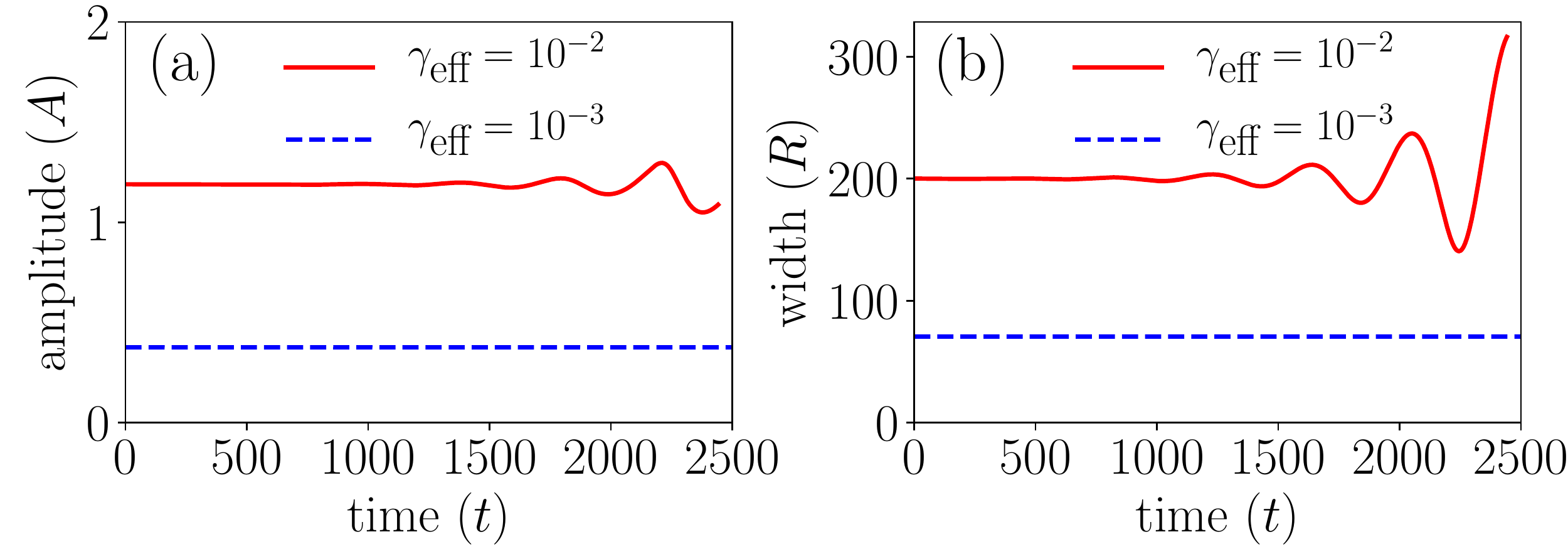}
  \caption{(Color online) Plots showing the time evolution of (a) amplitude ($A$) and (c) width ($R$) in the stable and unstable regimes with $\Omega = 0.01$, $g = 5$, $g_1 = 10^{-2}$, and with (i) $\gamma_{\text{eff}} = 10^{-3}$ (stable - dashed-blue line) and (ii) $\gamma_{\text{eff}} = 10^{-2}$ (unstable - solid-red line).}
  \label{fig:width}
\end{figure}%
\begin{figure}[!ht]
\centering
\includegraphics[width=\linewidth]{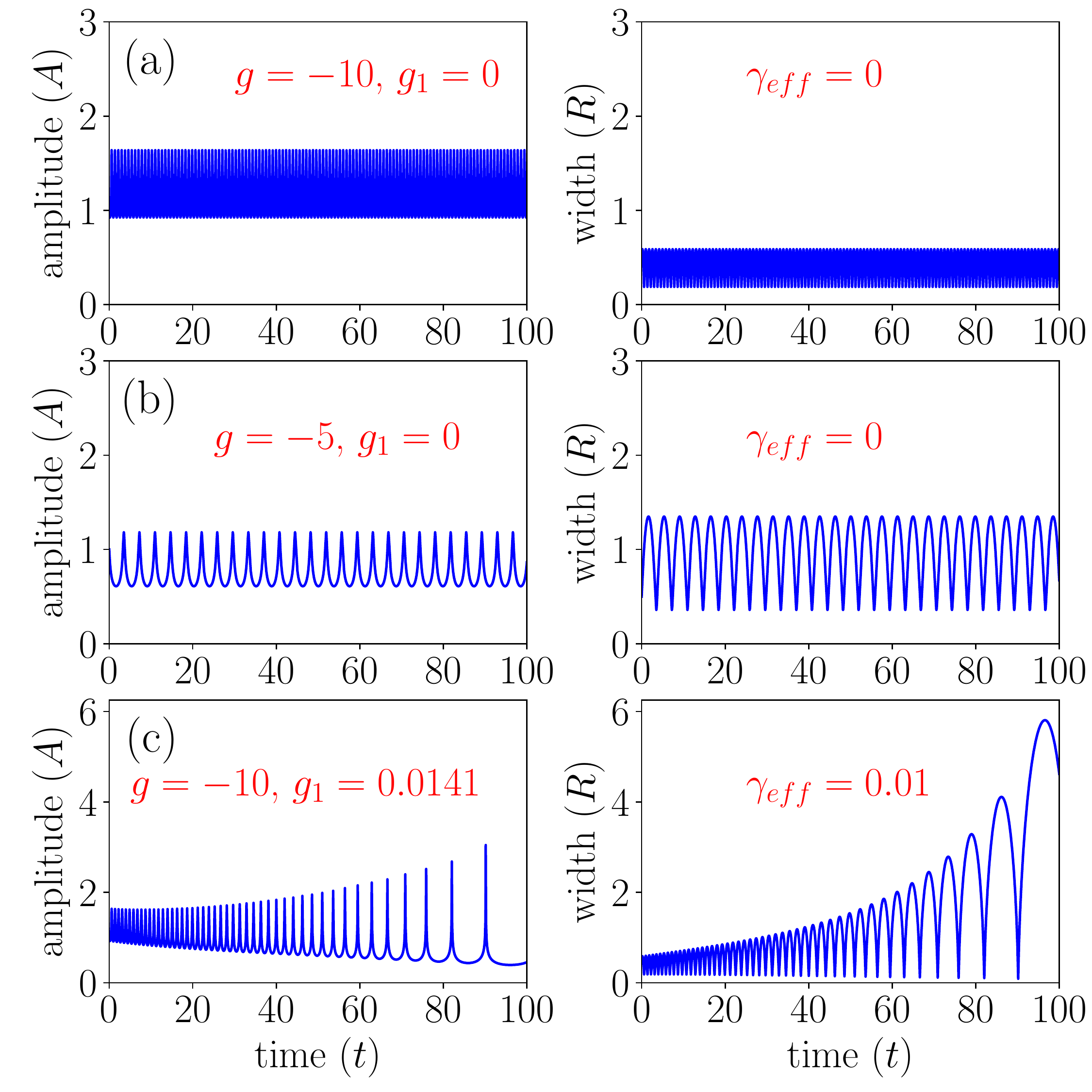}
\caption{(Color online) Amplitude \((A)\) and corresponding width \((R)\) of the trapless ($d(t) = 0$) attractive pBECs are shown in left and right panels, respectively. (a) $g=-10.0$, $g_1=0.0$ and $\gamma_{\text{eff}}=0.0$, (b) $g=-5.0$, $g_1=0$ and $\gamma_{\text{eff}}=0$, and (c) $g=-10.0$, $g_1=0.0142$ and $\gamma_{\text{eff}}=0.01$.}
\label{rf1}
\end{figure}%
It is pretty obvious from Fig.~\ref{fig:eig} that the area of the stability domain decreases as the trap gets weaker and weaker. It is also interesting to note that an unstable equilibrium point gets converted into a stable equilibrium point as the trap strength gets reduced. For example, $(g_1, \gamma_{\text{eff}}) = (10^{-2},10^{-4})$ which represents an unstable equilibrium point in Fig.~\ref{fig:eig}(a) for trap strength $\omega= 0.25$ attains stability in  Fig.~\ref{fig:eig}(d) at $\omega=0.001$. In addition, the maximum value of nonlinear  loss/gain $(g_1)$ representing a stable equilibrium point remains constant and is independent of the trap strength. We  also verify the stability regimes of the equilibrium point by numerically solving the above coupled system of equations given by (\ref{eq:odes2:final}) using the fourth order Runge-Kutta method with initial conditions chosen in the neighbourhood of the equilibrium point for further confirmation. Figs.~\ref{fig:width}(a) and \ref{fig:width}(b) depict the time evolution of the amplitude ($A$) and the corresponding  width ($R$) in the unstable  and stable regimes of the phase diagram shown in Fig.~\ref{fig:eig}(c) with $\Omega = 0.01$, $g=5$, $g_1 = 10^{-2}$ and $\gamma_{\text{eff}} = 10^{-3}$ (stable equilibrium) and $\gamma_{\text{eff}} = 10^{-2}$ (unstable equilibrium). The sustenance of  the amplitude and width of the condensates  for a longer interval of time is a clear signature of  their stability in the repulsive regime.
Next, to analyse the stability of the pBECs in detail, we solve Eqs.~\eqref{eq:odes:final} through the fourth order Runge-Kutta method and the results are reported below.

\subsection{Attractive pBEC}

In this section, by solving the Eqs.~\eqref{eq:odes:final:a} - \eqref{eq:odes:final:c}, we study the stability of the {\it trapless attractive} pBEC. 

In Fig.~\ref{rf1}, we show the dynamics of the two variational parameters, amplitude \((A)\) and corresponding width \((R)\) of the pBECs  shown in left and right panels, respectively. %
First, we begin with the following set of parameters to study the dynamics of the pBEc, $g=-10.0$, $g_1=0$, $\gamma_{\text{eff}}=0$ and $d(t) = 0$, shown in panel (a). The amplitude of the condensates start to stabilize once we ramp down the interaction strength from $g=-10$ to $g=-5$ as shown in panel (b). One also observes that the width of the condensates remains constant. When we ramp up the interaction strength again from $g=-5$ to $g=-10$  and  introduce  linear and nonlinear loss/gain terms, one witnesses small fluctuations in the amplitude and width of the condensates as shown in panel (c) indicating that the reinforcement of the cubic nonlinearity with nonlinear loss/gain terms in the attractive domain will render the condensates unstable. %
\begin{figure}[!ht]
\centering
\includegraphics[width=\linewidth]{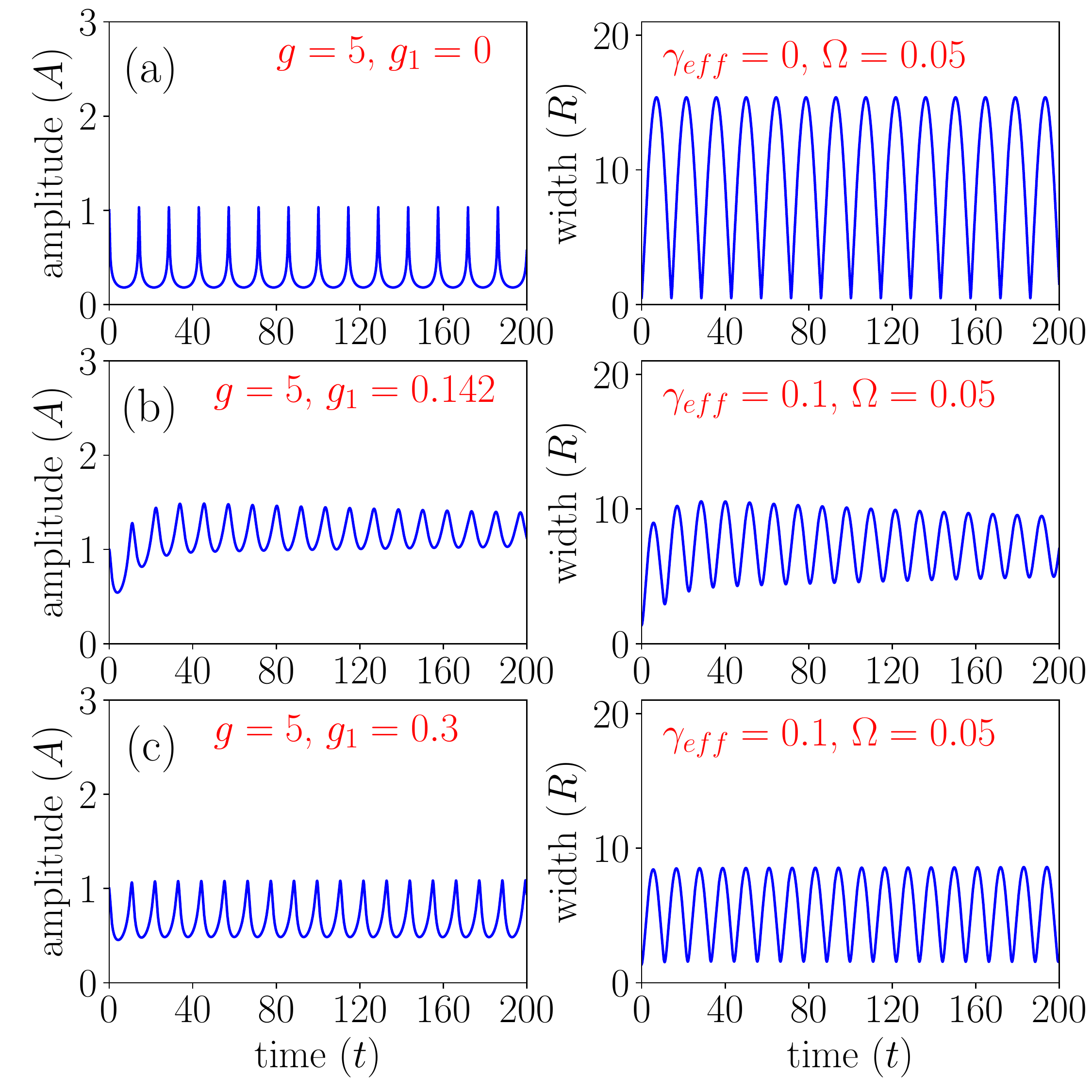}
\caption{(Color online) Amplitude \((A)\) and corresponding width \((R)\) of the repulsive pBECs with weak trap, $\Omega = 0.01$ and $d(t) = 1$, are shown in left and right panels, respectively: (a) $g=5.0$, $g_1=0.0$ and $\gamma_{\text{eff}}=0.0$, (b) $g=5.0$, $g_1=0.0$ and $\gamma_{\text{eff}}=0.1$, and (c) $g=5.0$, $g_1=0.3$ and $\gamma_{\text{eff}}=0.1$.}
\label{rf2}
\end{figure}

\subsection{Repulsive pBEC}

Next, we study the stability of the repulsive pBEC in the  presence of weak trap, $\Omega = 0.01$ and $d(t) = 1$. In Fig.~\ref{rf2}, we show the dynamics of the two variational parameters, amplitude \((A)\) and corresponding width \((R)\) of the repulsive pBECs shown in left and right panels, respectively. First, we begin with the following set of parameters to study the dynamics of the pBEc, \(g=5.0\), \(g_1=0\) and \(\gamma_{\text{eff}}=0\), shown in the first row [cf. panel (a) in Fig.~\ref{rf2}]. %
The amplitude \((A)\) and the corresponding width of the \((R)\) of the condensate are found to be smooth up to time \(t=200\). In the second row [cf. panel (b) in Fig.~\ref{rf2}], small variation starts to appear in the  amplitude \((A)\) and corresponding width \((R)\) of the condensate for \(g=5.0\), \(g_1=0.142\) and \(\gamma_{\text{eff}}=0.1\). However, this instability can be overcome in the  bottom row by manipulating the nonlinear loss/gain term and  one witnesses  both the  amplitude \((A)\) and the corresponding width \((R)\) of the condensate starting to stabilize for \(g=5.0\), \(g_1=0.3\) and \(\gamma_{\text{eff}}=0.1\) as shown in third row [cf. panel (b) in Fig.~\ref{rf2}]. It is pretty obvious from the dynamics of the above analysis that it is indeed possible to stabilize the condensates in the repulsive domain in the presence of a weak trap while the quantum of instability witnessed in attractive pBECs as shown in Fig~\ref{rf1}(c) is inevitable and this instability can be attributed to the reinforcement of nonlinear loss/gain term with cubic nonlinearity in the realm of nonequilibrium systems.  

\section{Numerical results}
\begin{figure}[!ht]
\begin{center}
\includegraphics[width=\linewidth]{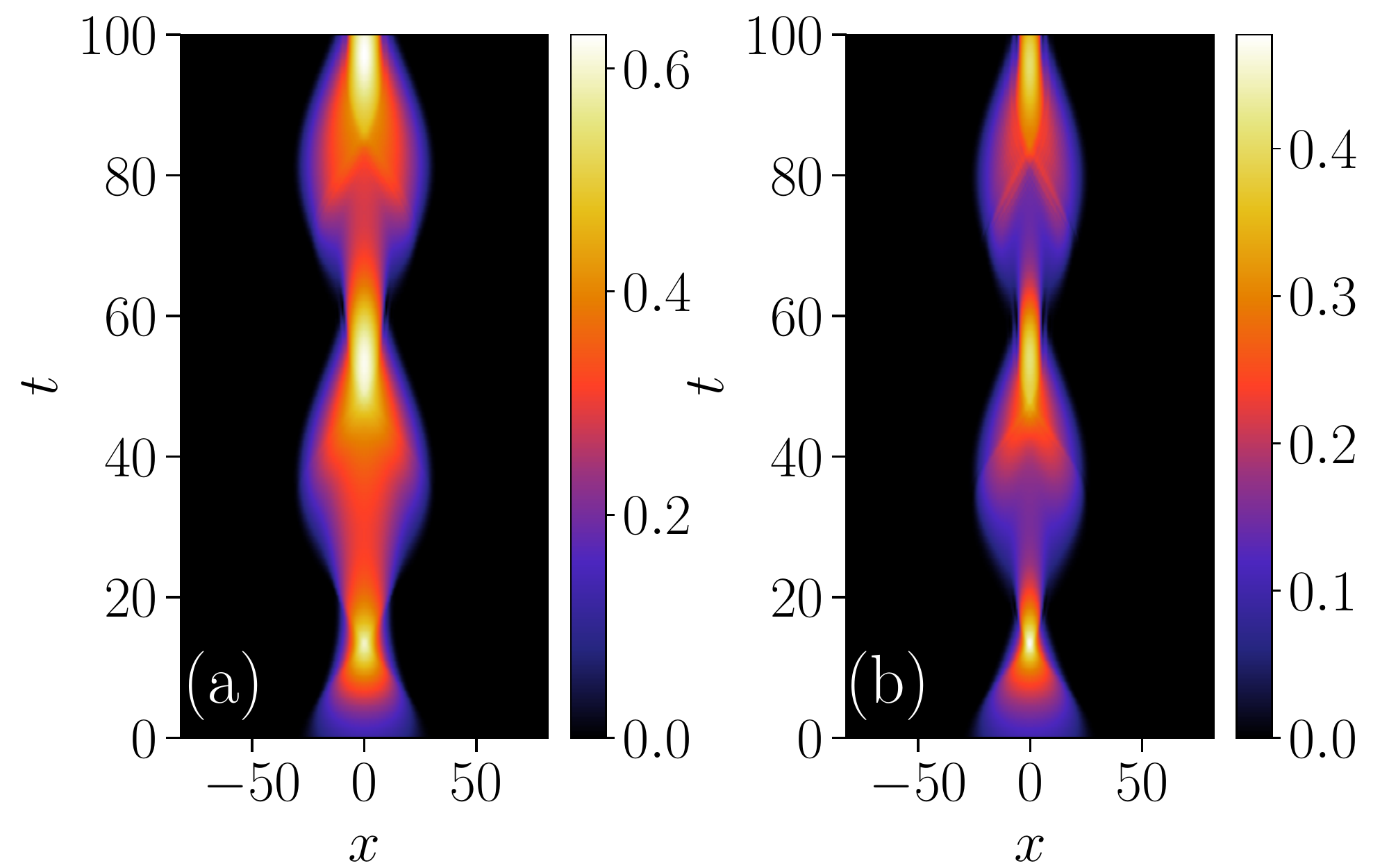}
\end{center}
\caption{(Color online) Density profiles for repulsive case for (a) $g=5.0$, $g_1=0.3$ and $\gamma_{\text{eff}}=0.1$, and (b) $g=5.0$, $g_1=0.5$ and $\gamma_{\text{eff}}=0.1$.} 
\label{rf8}
\end{figure}

In this section, we solve the time-dependent GP Eq.~(\ref{equ3}) numerically through the split-step Crank-Nicholson method~\cite{Muruganandam2009, Kumar2015}. In the course of the time evolution of the GP equation, certain initial conditions are necessary to stabilize a trapless system with a specific nonlinearity above a critical value. %
\begin{figure}[!ht]
\begin{center}
\includegraphics[width=\linewidth]{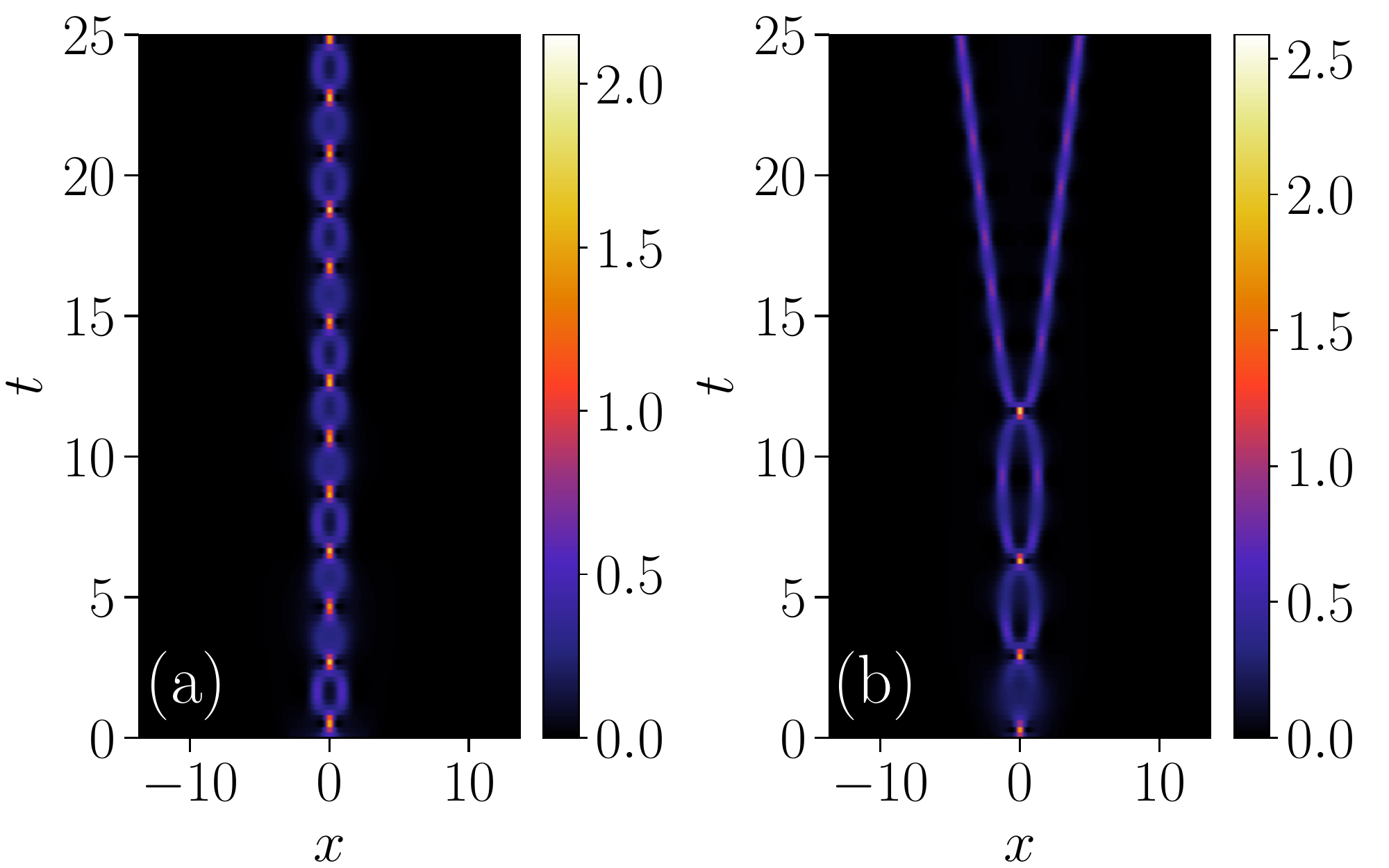}
\end{center}
\caption{(Color online) Density profiles for attractive case for (a) $g=-10.0$, $g_1=0.0$ and $\gamma_{\text{eff}}=0.0$, and (b) $g=-10.0$, $g_1=0.0142$ and $\gamma_{\text{eff}}=0.01$.}
\label{rf3}
\end{figure}%
If we choose the size of the system close to the desired size, then the system gets stabilized after a finite length of time. %
\begin{figure}[!ht]
\begin{center}
\includegraphics[width=\linewidth]{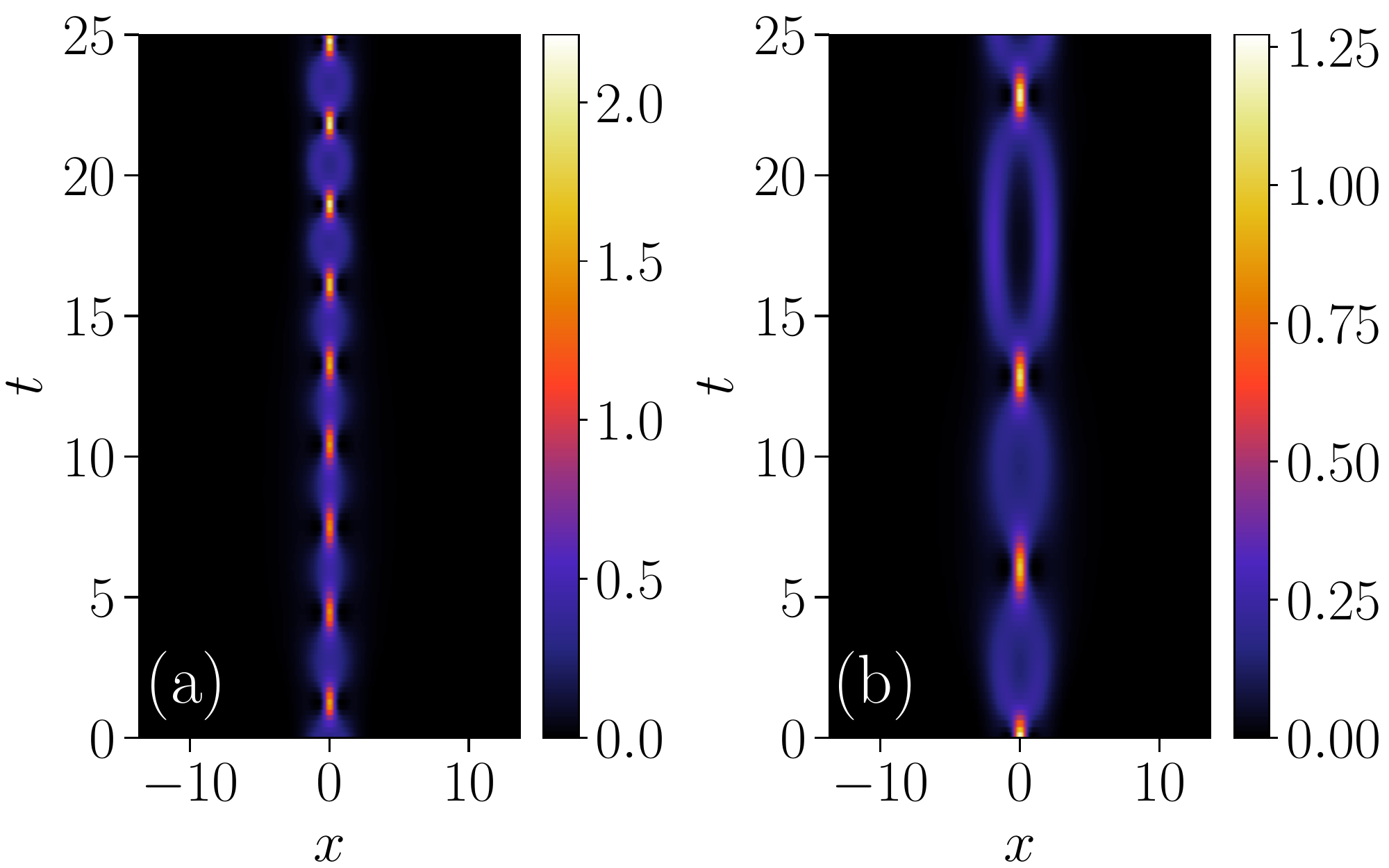}
\end{center}
\caption{(Color online) Density profiles for attractive case for (a) $g=-5.0$, $g_1=0.01$ and $\gamma_{\text{eff}}=0.01$, and (b) $g=-5.0$, $g_1=0.03$ and $\gamma_{\text{eff}}=0.01$.}
\label{rf4}
\end{figure}%
This procedure could be employed in an experimental setup to stabilize a trapless system. In the numerical simulation, it is necessary to remove the external trap while increasing the nonlinearity for obtaining stability. %
\begin{figure}[!ht]
\begin{center}
\includegraphics[width=\linewidth]{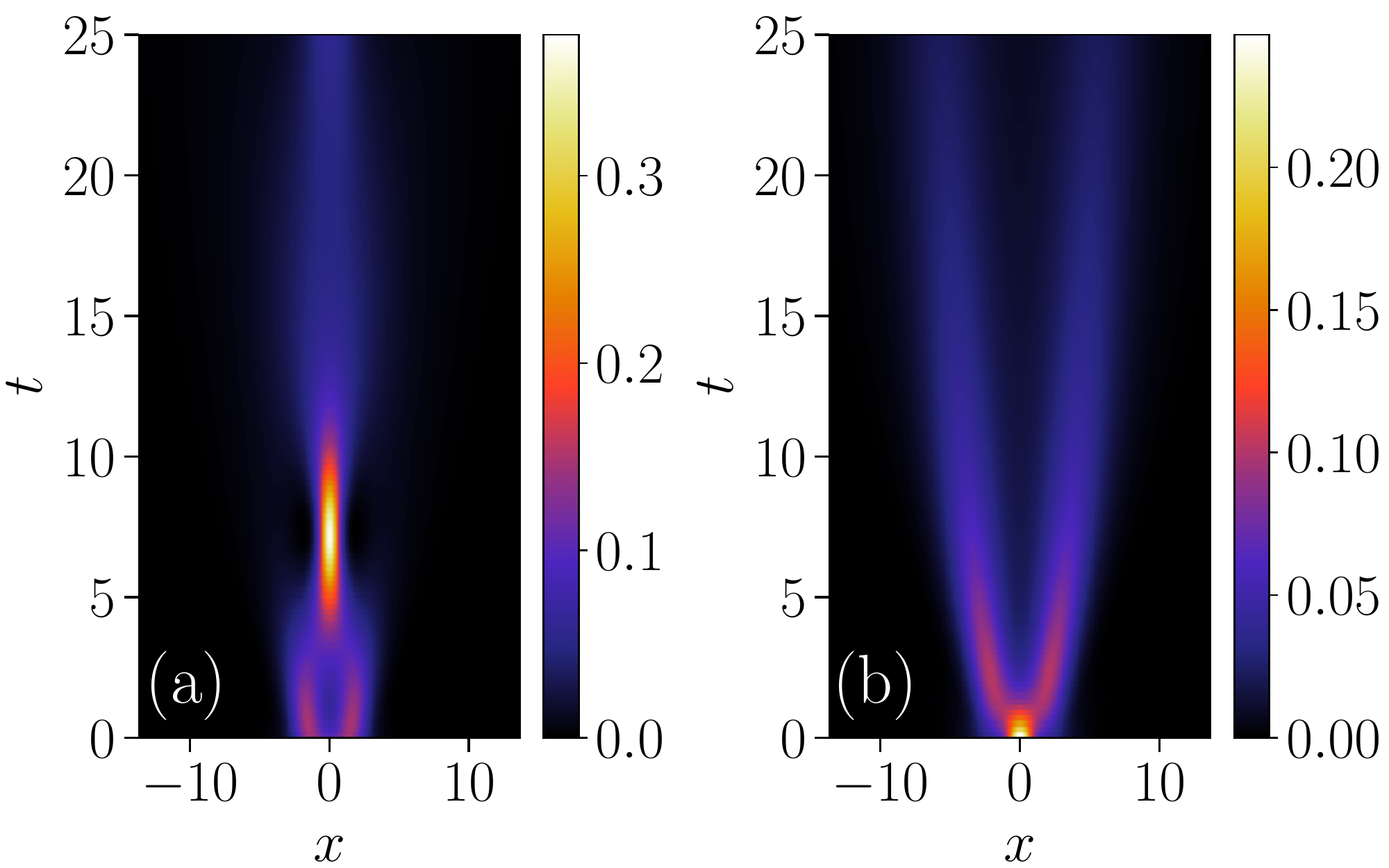}
\end{center}
\caption{(Color online) Density profiles for attractive case for (a)  $g=-5.0$, $g_1=0.03$ and $\gamma_{\text{eff}}=-0.01$, and (b) $g=-5.0$, $g_1=0.1$ and $\gamma_{\text{eff}} = -0.01$.}
\label{rf5}
\end{figure}%
In the course of the time iteration, the coefficients of the nonlinear terms are increased from $0$ at each time step as $g(t)= f(t) g$, where $f(t)=t/\tau$ for $0 \leq t \leq \tau$ and $f(t)=1$ for $t \geq \tau$. At each time, the trap is removed by changing $d(t)$ from $1$ to $0$ by $d(t) = 1 -f(t)$. During this process, the harmonic trap is removed and the nonlinearity $g$ attained at time $\tau$. %
\begin{figure}[!ht]
\begin{center}
\includegraphics[width=\linewidth]{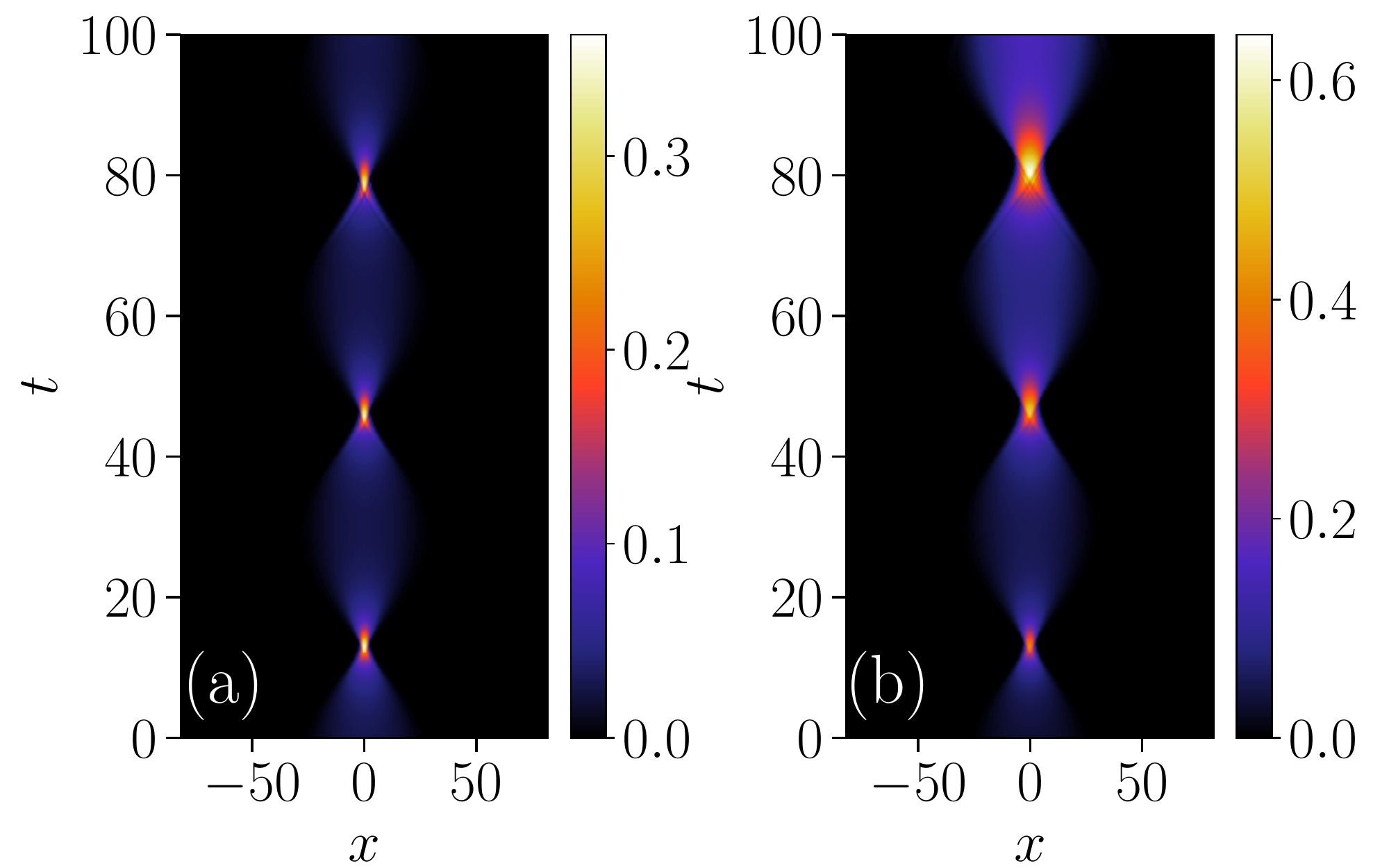}
\end{center}
\caption{(Color online) Density profiles for repulsive case for (a) $g=5.0$, $g_1=0.0$ and $\gamma_{\text{eff}}=0.0$, and (b) $g=5.0$, $g_1=0.05$ and $\gamma_{\text{eff}}=0.05$.} 
\label{rf6}
\end{figure}%
The density profiles for attractive condensates for $g=-10$, $g_1=0$ and $\gamma_{\text{eff}}=0$ is shown in Fig. \ref{rf3}(a). In Fig.~\ref{rf3}(a), one witnesses the ``beating effect'' which corresponds to the linear superposition of the condensates at periodic intervals of time. %
\begin{figure}[!ht]
\begin{center}
\includegraphics[width=\linewidth]{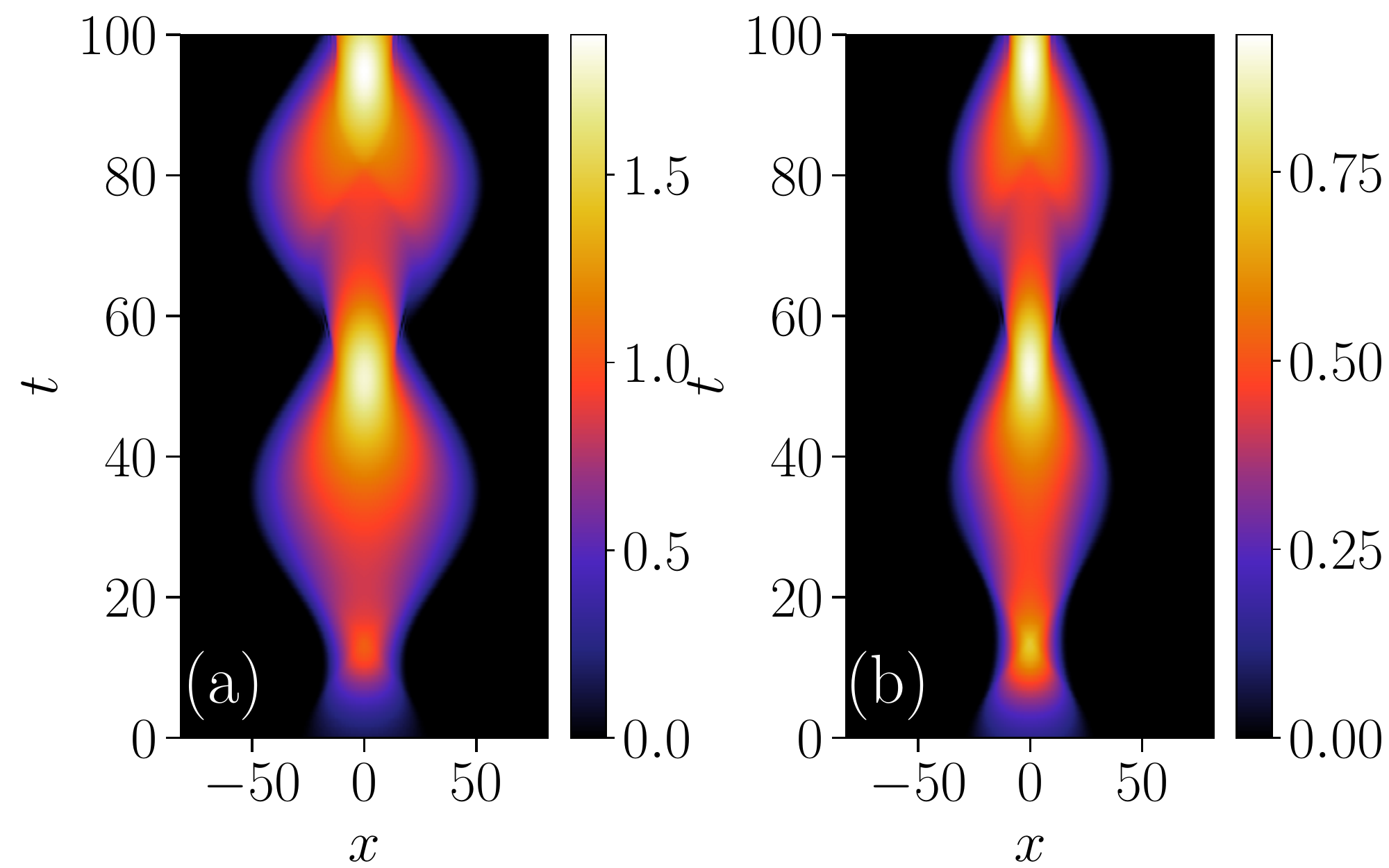}
\end{center}
\caption{(Color online) Density profiles for repulsive case for (a) $g=5.0$, $g_1=0.1$ and $\gamma_{\text{eff}}=0.1$, and (b) $g=5.0$, $g_1=0.2$ and $\gamma_{\text{eff}}=0.1$.} 
\label{rf7}
\end{figure}%
We also find that the periodicity of the beating effect can be manipulated by the simultaneous introduction of linear and nonlinear loss/gain terms as shown in Figs.~\ref{rf3}(b), \ref{rf4} and \ref{rf5}. On the other hand, the density profiles of the condensates with repulsive interactions with weak trap are shown in Figs.~\ref{rf6}, \ref{rf7} and \ref{rf8}. From the figs,  one again observes the beating effect. But, unlike the attractive pBECs, one witnesses the stretching of the density of the condensates when we increase the strength of nonlinear loss/gain terms. In addition, the stretching occurs periodically and the density pattern looks like a classical stretched string with nodes and antinodes.

\section{Conclusion}

We have theoretically studied the stability properties of trapless polariton BECs through a judicious interplay between two-body interaction, linear and nonlinear loss/gain. We have then solved the complex Gross-Pitaevskii equation through the variational approach and obtained the ordinary differential equations for the amplitude and width of the condensate. These ordinary differential equations are then rewritten to perform a linear stability analysis to generate a stability window in the repulsive domain. We have then supported our variational results by numerical simulations through Runge-Kutta method which are then reaffirmed by split-step Crank-Nicholson method. %
The density profiles of attractive pBECs exhibit ``beating effects'' wherein the condensates superpose at periodic intervals of time and the periodicity can be manipulated by the introduction of linear and nonlinear loss/gain terms. For repulsive pBECs, one witnesses the stretching of the density while the periodicity of the loop remains a constant. We do believe that this simple protocol, to identify the possible stability windows of pBECs particularly in the repulsive regime will open up a lot of avenues in future from an experimental perspective.


\acknowledgments
Authors are indebted to the referees for their critical comments and thinking  which have immensely improved the results of the paper. SS acknowledges partial support from the Council of Scientific and Industrial Research (CSIR) for financial assistance (Grant No. 03(1456)/19/EMR-II) and the Foundation for Research Support of the State of São Paulo (FAPESP) [Contracts 2020/02185-1 and 2017/05660-0]. RKK acknowledges support from Marsden Fund (Contract UOO1726). 
RR wishes to acknowledge the financial assistance from DAE-NBHM (Grant No. 02011/3/20/2020/NBHM(R.P)/RD II) and CSIR (Grant No 03(1456)/19/EMR-II).
The work of PM is supported by DST-SERB (Department of Science \& Technology - Science and Engineering Research Board) under Grant No. CRG/2019/004059. 


%

\end{document}